# High pressure mediated physical properties of Hf$_2$AB (A = Pb, Bi) via DFT calculations

M. S. Hossain[1], N. Jahan[1], M. M. Hossain[1,2], M. M. Uddin[1,2], M. A. Ali[1,2]*

[1]Department of Physics, Chittagong University of Engineering and Technology (CUET), Chattogram-4349, Bangladesh
[2]Advanced Computational Materials Research Laboratory (ACMRL), Department of Physics, Chittagong University of Engineering and Technology (CUET), Chattogram-4349, Bangladesh

## ABSTRACT

Using density functional theory (DFT), the structural, mechanical, electronic, thermal, and optical properties of Hf$_2$AB (A = Pb, Bi) borides were studied, considering the pressure effect up to 50 GPa. The lattice constants were found to be decreased with increasing pressure wherein the lattice constants at 0 GPa agree well with the reported values. The stability (mechanical and dynamical) of the titled compounds at different pressures was checked. The mechanical behavior was disclosed considering the bulk modulus, shear modulus, Young's modulus, Pugh ratio, Poisson's ratio, and hardness parameter at different pressures. Pugh's and Poisson's ratios were used to assess the brittleness and ductility of the titled borides. The anisotropic nature of mechanical properties was studied by calculating different indices and plotting 2D and 3D projections of the elastic moduli. The electronic properties were revealed by calculating the band structure, density of states, and charge density mapping at different pressures, wherein the anisotropic nature of the electronic conductivity was noted. We studied the Debye temperature, minimum thermal conductivity, Grüneisen parameter, and melting temperature of the titled borides at different pressures; the results revealed the improvement of the mentioned properties with rising pressure. The important optical constants to disclose the possible relevance in application purposes were investigated; a little pressure effect was noted. The thermal properties suggest that the titled borides could be used as thermal barrier coating (TBC) materials while the reflectivity spectra revealed their suitability to be used as cover materials for protection from solar heating.

**Keywords**: High pressure; Hf$_2$AB (A = Pb, Bi); mechanical behavior; electronic properties; thermal properties; optical properties.

Corresponding author: ashrafphy31@cuet.ac.bd

## 1. Introduction

The general formula for MAX Phase is $M_{n+1}AX_n$, where n is a defined numerical number, M indicates an early transition metal, A comes from A-group elements, and X will be C/N/B or a mixer of them. Interestingly, the MAX Phase materials consist of a novel mixture of both ceramics and metallic characters [1]. The researchers have given their attention because of their metallic characteristics, such as mechanical strength, conductivity, and machinability. Whereas their behavior is also fine at high-temperature, corrosion and oxidation resistance are also good as ceramics compounds [2-5]. These lucrative characteristics persuade the scientific community to explore and synthesize the MAX phases for possible relevance of practical usefulness in many sectors [2]. Many researchers have tried to explore these materials theoretically or experimentally. Theoretically, many compounds have already been explored but experimentally not synthesized yet. More than 80 MAX phase members have already been synthesized and have become classified as a large family with 150 plus known MAX phase members [6] out of 665 possible MAX phases [7]. MAX phases are classified into three sub-groups based on the value of n: $M_2AX$ is known as 211 phase (n = 1), $M_3AX_2$ is 312 (n = 2) phase, and finally $M_4AX_3$ is 413 phase (n= 3). The 211 phases are more in number than the other phases. In addition, a combination of 211, 312, and 413 phases can form some of the hybrid MAX phases [8].

In the 1960s, many 211 (such as $Ti_2AlC$) and two 312 (e.g., $Ti_3SiC_2$ and $Ti_3GeC_2$) phases were explored by Nowotny and his colleagues [9]. Twenty years later, his (Nowotny) student, Schuster, discovered $Ti_3AlC_2$ as a member of 312 MAX phases [10]. But, the researchers did not show their interest in the MAX phase before the legendary work on $Ti_3SiC_2$ published by Barsoum and El-Raghy; they have discovered the unique character of $Ti_3SiC_2$, which exhibits both metallic and ceramics properties [11, 12]. The compound $Ti_4AlN_3$ had included as a first member of 413 (n = 3) phases in 1999 [13, 14]. Later, Palmquist et al. [15] explored higher-ordered MAX phases such as 514, 615, and 716. The hybrid phases consist of 211, 312 and 413 phases; for example, the unit cell of the 523 phase is composed of 312 and 211 phases. The 725 phase comprises 312 and 413 phases [16]. The researchers also discovered a new class of 2D materials called MXene in the 2010s [17, 18]. The synthesis of MXene is usually done by selective etching of A elements from their MAX phases. The MXenes are considered promising candidates to keep energy stored and used in many electronic devices. There are also a considerable variety of MAX phase solid solutions with different combinations of M, A, or X site elements [19-24].

Although the MAX phase's members are grown up by tuning M or A elements, however, their diversity for the X element is limited up to C or N for a long time [25]. Thus, an extension of X elements can open a new door and add a huge number of possible new compounds to the MAX phase family. It is great news that the scientific community has already made an extension of the X (C/N) element up to the B atom [25, 26]. The B and related borides have already been proven as potential materials that can be used in many fields [27-29]. The extension has also been confirmed by successfully synthesizing some MAX phase borides [30-32]. Zhang et al. [30] have synthesized $Hf_2TeB$ boride. $M_2SB$ (M = Zr, Hf, and Nb) borides have been synthesized by Rackl *et al.* [31, 32]. The physical properties of $M_2SB$ (M = Zr, Hf, and Nb) borides were also studied later by Ali et al. [33] via first-principle calculations. The obtained mechanical, electronic, thermal, and optical properties of $M_2SB$ (M = Zr, Hf, and Nb) borides confirmed that they are also potential candidates for use in place of earlier MAX phase carbides. Wang *et al.* [34] discovered another 212 MAX phase boride ($Ti_2InB_2$), further studied by Ali *et al.*[35] and Wang *et al.*[36]. Moreover, attempts were also made to explore MAX phase borides by some other researchers. For example, a series of MAX phase borides [$M_2AlB$ (M = Sc, Ti, Cr, Zr, Nb, Mo, Hf, or Ta)] were predicted by Khazaei *et al.*[25]. They investigated wherein the formation energy and electronic and mechanical properties. Later on, the $Ti_2SiB$ [37], $M_2AB$ (M = Ti, Zr, Hf; A = Al, Ga, In) [38], $V_2AlB$ [39], and $M_2AlB$ (M = V, Nb, Ta) [40] phases were also been predicted to be stable. In all cases, the new MAX phase borides have been proven as an alternative to those of earlier MAX phase carbides. Miao *et al.* [41] predicted some unconventional MAX phase borides that are crystallized in slightly different space groups (SG: $P\bar{6}m2$, 187) of hexagonal system, a sub-space group of MAX phase as defined by them. In addition, Miao *et al.*[41] have also predicted some 211 MAX phase borides based on the thermo-dynamical stability. From the predicted borides, $Hf_2AB$ (A = Pb, Bi) borides are selected for the present study.

In this paper, the effect of pressure on the physical properties of titled borides is considered, motivated by recent studies assessing the pressure effect on other MAX phase materials [3, 5, 42-51]. The pressure has a tremendous impact on the properties owing to the shrinkage of the unit cell; consequently, it decreases the inter-atomic distance and enhances the bonding strength among the atoms. For example, the materials exhibit a transition from brittle to ductile. The mechanical and thermal properties significantly increased with rising pressure.

In the present research, the applied pressure significantly improves the mechanical and thermal properties. However, the minimum thermal conductivity values increased with pressure but remained reasonably suitable for high-temperature applications such as thermal barrier coating materials. We hope this article will provide some vital information about the titled borides, which might be helpful for the selection of suitable materials for practical applications. Therefore, the effect of pressure on the structural, electronic, mechanical, thermal, and optical properties of $Hf_2AB$ (A = Pb, Bi) borides is presented in this paper.

## 2. Computational methodology

The properties of our titled borides have been calculated by plane-wave pseudopotential that works following density functional theory based on quantum mechanics. All the process is implemented on CASTEP (CAmbridge Serial Total Energy Package) code [52, 53]. GGA has been used in Perdew–Burke–Ernzerhof (PBE) [54] for measuring exchange-correlations energy. For optimizing energy, k-point mesh [55] has been taken 500 eV, $10 \times 10 \times 3$, respectively. Atomic configuration is set by BFGS [56], and density mixing is used to optimize the electronic structure. Others parameters setting values are $5 \times 10^{-6}$ eV/atom for energy convergence tolerance, and 0.01 eV/Å has been used for maximum force tolerance. Maximum ionic displacement and stress tolerances are $5 \times 10^{-4}$ Å and 0.02 GPa, respectively.

## 3. Results and discussion

### 3.1. Structural properties

The compound $Hf_2AB$ (A = Pb, Bi) is a hexagonal structure with a space group of $P6_3/mmc$, as shown in Fig. 1 (a). There are three elements in this compound, namely: Hf, Pb/Bi, and B and their atomic positions are (1/3, 2/3, $Z_M$), (1/3, 2/3, 3/4), and (0, 0, 0), respectively. After energy minimization, the lattice constant $a$ and $c$, and the internal parameters $Z_M$ were obtained. The geometrical optimization of the structures was performed at different pressures to understand the behavior of $Hf_2AB$ (A = Pb, Bi) under pressure, and the equilibrium lattice constants are presented in Table 1. As seen, the lattice constants are noted to be decreased with increasing pressure. Figs. 2(a, b, c) displays the normalized lattice constants ($a/a_0$, $c/c_0$) and volume ($V$) vs pressure graphs to visualize the pressure effect. A gradual decrease of $a/a_0$, $c/c_0$, and $V$ with the rise in pressure is seen from Figs. 2(a, b, c). The inter-atomic distance of the lattice decreases under hydrostatic pressure resulting in the decrease of these parameters. As there is no breaking point in the decreasing graphs, thus, no structural phase transition is expected to occur in these compounds within the studied pressure range. Table 1 also lists the $c/a$ ratio, an important parameter to exhibit the anisotropy of a hexagonal system, and it is usually 4 for 211 MAX phases [57]. Thus, the lattice constant $c$ is usually four times higher than $a$; the atomic arrangement is also different along two axes ($a$ and $c$). As evident from Fig. 1, the atomic density along the $c$-direction is lower than that of the $a$-direction. Thus, the compressibility along the $c$-axis is expected to be lower than that of the $a$-axis.

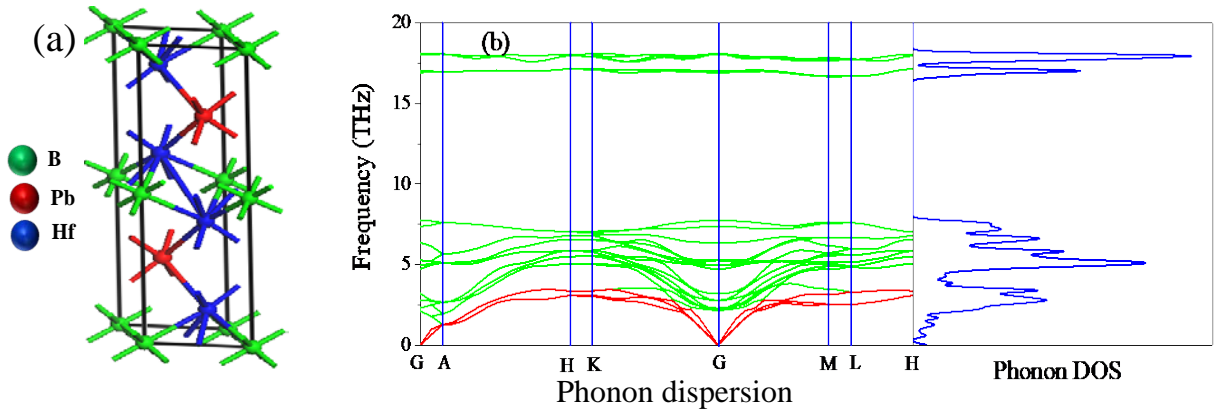

**Fig. 1: (a)** Crystal structure (unit cell) and (b) phonon dispersion curve (PDC) and phonon DOS (PHDOS) of Hf$_2$PbB.

Within the pressure range 0-50 GPa, the decrease of $a/a_0$ is found to be 8.16 % for Hf$_2$PbB and 8.44% for Hf$_2$BiB from 0 GPa to 50 GPa while it is found to be 7.82 and 6.54 % for Hf$_2$PbB and Hf$_2$BiB borides, respectively for $c/c_0$ ratio. The $V/V_0$ ratio decreases by 22.27% and 21.64% for Hf$_2$PbB and compounds, respectively, for the same pressures. The value of $c/a$ ratio with pressure varies for both compounds. For, it decreases up to 20 GPa, after then increases with pressure, while it increases gradually for Hf$_2$BiB with pressure. Though both $a$ and $c$ decreases due to hydrostatic pressure, the decrease in $a$ is lesser up to 20 GPa for Hf$_2$PbB; consequently, the $c/a$ ration decreases for the same. As seen, the decrease of $a/a_0$ ($c/c_0$) is found to be 8.16 (7.82) % for Hf$_2$PbB while it is 8.44 (6.54) % for Hf$_2$BiB from 0 GPa to 50 GPa. The distortion of the crystal has been calculated using the internal parameter $Z_M$ and $c/a$ ratio. The MAX phase with hexagonal space group (SG: 194) is formed by the M$_6$X octahedron and M$_3$A trigonal prism. If the octahedron and trigonal parameters are equal to 1, the polyhedra will be considered ideal-deviation from 1 measures the distortion of the polyhedral. The smaller polyhedra distortion indicates the compounds' stability [58]. The distortion parameters of both compounds are calculated using the optimized values. It is seen from Fig. 2(d) that $O_r$ for both compounds decrease with the increasing pressure, *i.e.,* the compounds get more stable, but the trigonal parameter ($P_r$) increases for the same. At 0 GPa pressure, the octahedron and trigonal prism ratio are 1.03 and 1 for the Hf$_2$PbB and Hf$_2$BiB borides, respectively. The value of $O_r$ and $P_r$ are 1.088 and 1.081, reported for the Ti$_2$GaC MAX phase [59]. The values ($O_r$ and $P_r$) for the studied compounds are lower than those of the referenced phase. Fig. 1 (b) also shows the phonon dispersion curve and corresponding DOS of Hf$_2$PbB, wherein no imaginary frequency exists, confirming the dynamical stability of Hf$_2$PbB. The PDC and PHDOS of Hf$_2$AB (A = Pb, Bi) at different pressures (0 – 50 GPa; in a step of 10) have also been calculated [not shown to save the journal page] and also obtained similar curve for each phase and pressure without having any imaginary frequency. Thus, the studied compounds are said to be dynamically stable up to 50 GPa, at least.

**Table 1:** Calculated lattice parameters (*a* and *c*) in Å, unit cell volume V(Å$^3$), hexagonal ratio (*c/a*), internal parameter ($Z_m$), distortion parameters for octahedra ($O_r$) and trigonal prism ($P_r$) of Hf$_2$AB (A = Pb, Bi) at different pressures.

| Phase | Pressure (GPa) | *a* (Å) | *c* (Å) | *v* (Å$^3$) | *c/a* | $Z_M$ | $O_r$ | $P_r$ |
|---|---|---|---|---|---|---|---|---|
| | 00 | 3.493 | 14.896 | 157.462 | 4.265 | 0.0840 | 1.121 | 1.094 |
| **Hf$_2$PbB** | 05 | 3.448 | 14.674 | 151.049 | 4.256 | 0.0853 | 1.108 | 1.101 |
| | 10 | 3.409 | 14.499 | 145.901 | 4.253 | 0.0863 | 1.097 | 1.105 |
| | 15 | 3.374 | 14.351 | 141.496 | 4.253 | 0.0871 | 1.089 | 1.108 |
| | 20 | 3.345 | 14.224 | 137.794 | 4.252 | 0.0878 | 1.081 | 1.111 |
| | 25 | 3.316 | 14.110 | 134.497 | 4.255 | 0.0883 | 1.075 | 1.113 |
| | 30 | 3.292 | 14.028 | 131.629 | 4.261 | 0.0887 | 1.070 | 1.114 |
| | 35 | 3.268 | 13.942 | 128.985 | 4.266 | 0.0891 | 1.064 | 1.114 |
| | 40 | 3.247 | 13.865 | 126.598 | 4.270 | 0.0895 | 1.059 | 1.115 |
| | 45 | 3.227 | 13.795 | 124.411 | 4.275 | 0.0898 | 1.055 | 1.116 |

| | | | | | | | |
|---|---|---|---|---|---|---|---|
| | 50 | 3.208 | 13.731 | 122.393 | 4.280 | 0.0900 | 1.052 | 1.116 |
| **Hf$_2$BiB** | 00 | 3.530 | 14.374 | 155.123 | 4.072 | 0.0867 | 1.135 | 1.135 |
| | 05 | 3.483 | 14.197 | 149.178 | 4.076 | 0.0878 | 1.122 | 1.139 |
| | 10 | 3.443 | 14.061 | 144.342 | 4.084 | 0.0886 | 1.111 | 1.141 |
| | 15 | 3.407 | 13.940 | 140.124 | 4.092 | 0.0892 | 1.103 | 1.142 |
| | 20 | 3.375 | 13.840 | 136.563 | 4.100 | 0.0898 | 1.094 | 1.143 |
| | 25 | 3.346 | 13.757 | 133.375 | 4.111 | 0.0903 | 1.087 | 1.143 |
| | 30 | 3.318 | 13.692 | 130.558 | 4.126 | 0.0906 | 1.080 | 1.142 |
| | 35 | 3.294 | 13.624 | 128.013 | 4.136 | 0.0910 | 1.074 | 1.142 |
| | 40 | 3.273 | 13.545 | 125.686 | 4.138 | 0.0913 | 1.070 | 1.143 |
| | 45 | 3.253 | 13.488 | 123.576 | 4.147 | 0.0916 | 1.065 | 1.143 |
| | 50 | 3.232 | 13.434 | 121.545 | 4.156 | 0.0918 | 1.061 | 1.142 |

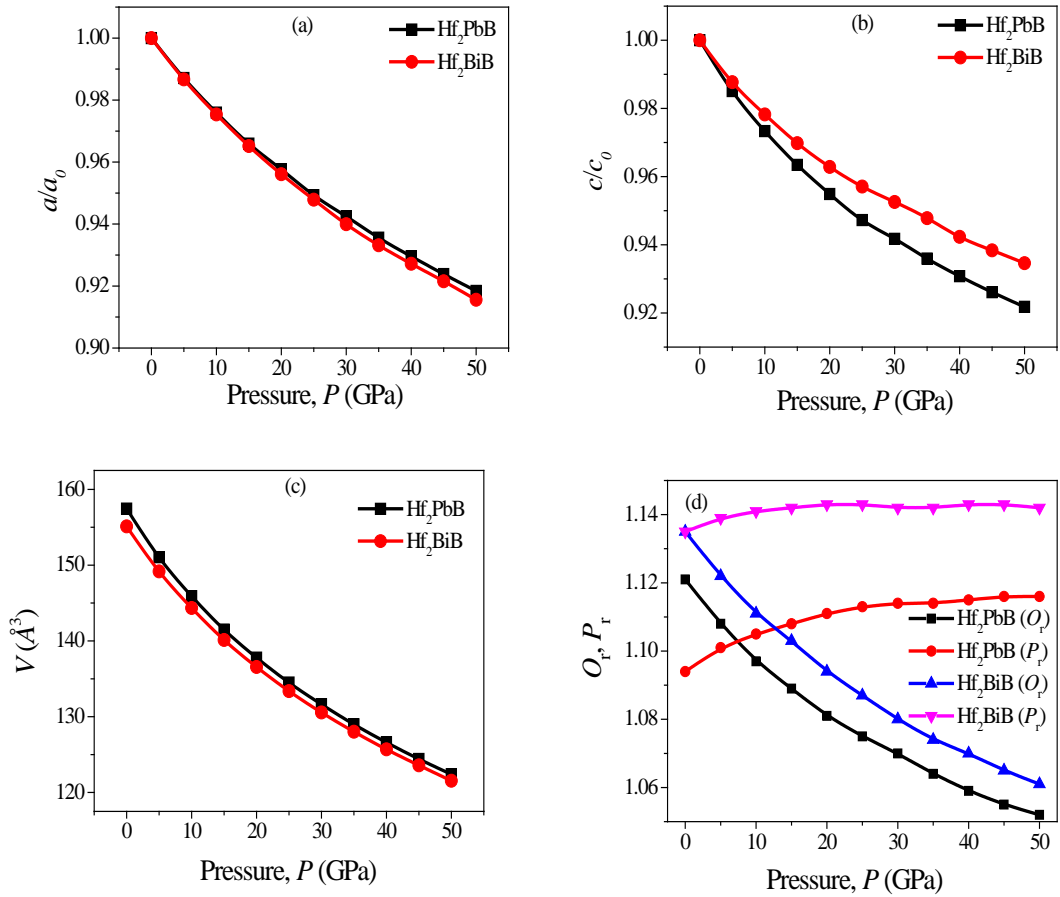

**Fig. 2:** Variation of normalized lattice parameters (a) $a/a_0$ and (b) $c/c_0$, (c) volume, and (d) distortion parameters: octahedron ($O_r$) and trigonal prism ($P_r$) as a function of pressure.

## 3.2. Electronic Properties

### 3.2.1 Density of states

The total and partial density of states (DOS) provides the bonding nature of materials. The calculated DOS of $Hf_2AB$ (A = Pb, Bi) is presented in Figs. 3(a-f) to discuss the electronic structure. The DOS for every pressure step has been calculated but only presented [total and partial DOS] for 0 GPa, 30 GPa and 50 GPa due to almost similar nature and to save the page. From these plots, one can see that the main contribution at the Fermi level comes from the Hf-5$d$ orbitals, while a little contribution from the Pb/Bi-6$p$ orbitals is also noticed. The valence band can be divided into two energy ranges such as lower valence band: (- 12.0 eV to -6.0 eV) and upper valence band (-6.0 eV to - 0.0 eV). In the energy range- 12.0 eV to -6.0 eV, Pb/Bi-6$s$ and B-2$s$ orbitals are hybridized, and the contribution of Pb/Bi-6$s$ states is more than that of B-2$s$ states to the peak. Both compounds show a similar nature, and the value of DOS decreases with an increase in pressure. At 0 GPa, the DOSs at the Fermi level are 3.68 and 3.29 states/eV/unit cell for the $Hf_2PbB$ and $Hf_2BiB$ phase, respectively. The DOSs in the pressure range 0-50 GPa in step 5 GPa have been depicted in Figs. 3(g, h) for both compounds. The slope in DOS reduces and spreads with an increase in pressure, the flatness of the DOS indicates the enhancement of the electrical conductivity [60].

### 3.2.2 Electronic band structure

Using GGA-PBE functional at equilibrium lattice structures, the band structure of the $Hf_2AB$ (A=Pb, Bi) MAX phase has been calculated and discussed at different pressures in the energy range from -15 to 6 eV, and the Fermi level is set at 0 eV. Fig. 4(a, b, c) shows the band structure of $Hf_2PbB$ at 0, 30, and 50 GPa, whereas Fig. 4(d, e, f) shows the same for $Hf_2BiB$. There is no band gap in these band structures owing to the overlapping valence and conduction bands. As a consequence, the compounds are said to be metallic.

Energy dispersion along the $c$ direction is shown by Γ-A, H-K, and M-L paths. On the other hand, A-H, K-Γ, Γ-M, and L-H paths lead the energy dispersion corresponding to the basal plane. Because of the higher effective mass tensor along the $c$ direction, the energy dispersions are lower than those of the basal plane. Consequently, the electrical conductivity of the compounds is said to be anisotropic. As seen from the Figures, energy bands of these compounds cover more energy range with increasing pressure. The bands move to the Fermi level with increased pressure, as shown inside the circle. As the hydrostatic pressure is increased, the effective electron's mass tensor becomes smaller, so the electrical conductivity is increased [60].

### 3.2.3 Charge density

The main issue of charge density distribution is to explore the chemical bonds present within the compounds corresponding to the electron densities. The charge density mapping has been calculated for different pressure up to 50 GPa with a step of 5 GPa. The red color region refers to a high electronic charge density, whereas the blue color refers to a low electronic charge density. One can understand the positive electron density regions from the charge accumulation. The charge depletion indicates the negative charge density region [61]. A stronger covalent bond can be formed by accumulating charge in the vicinity of atomic position, whereas negative charges are responsible for creating ionic bonds. Fig. 5(a-b) illustrates the charge density mapping for the Hf$_2$PbB and Hf$_2$BiB compounds, respectively. The figures have substantial strong charge accumulation at the position of Hf and B atoms for both compounds. The charge accumulation and depletion at their atomic region of Hf$_2$PbB compounds is slightly different from the Hf$_2$BiB phase, which implies that the bonding strength and the amount of charge transfer could be different for them. Mulliken population analysis confirms the before-said statement. Hf and B atoms form the covalent bond. Fig. 5(a-b) also demonstrates the evaluation of charge density with the rise in pressure. As evident, the bond strength of the studied compounds is observed to be increased with pressure which is in good agreement with the calculated values of the Vickers hardness presented in Table 3.

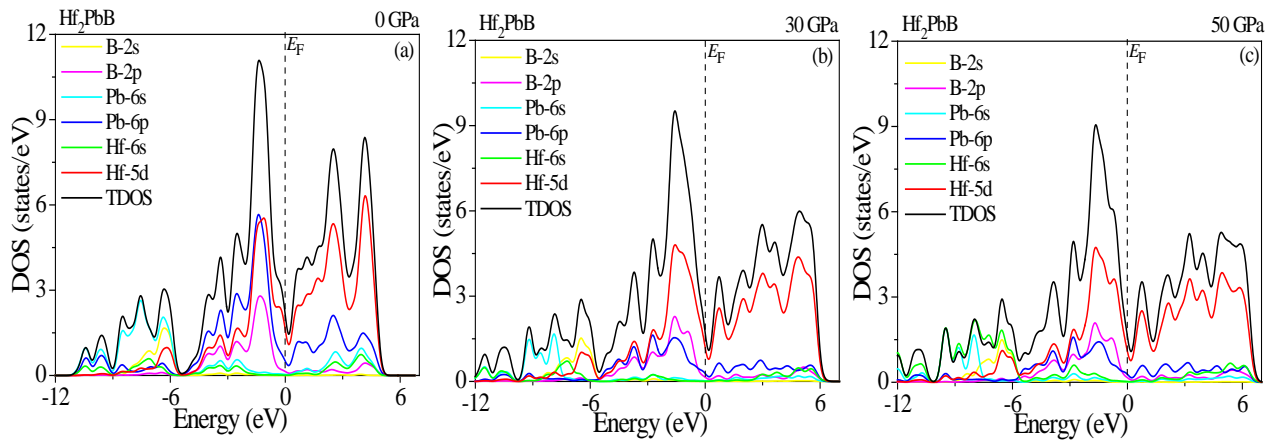

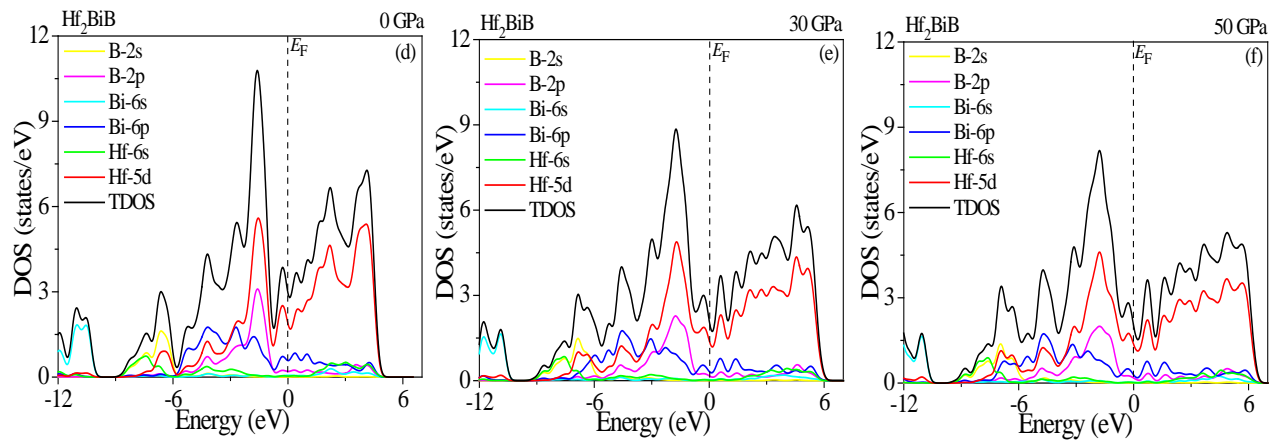

**Figs. 3:** Total and partial density of states of Hf₂PbB (a, b, c) and Hf₂BiB (d, e, f) at 0 GPa, 30 GPa, and 50 GPa pressure.

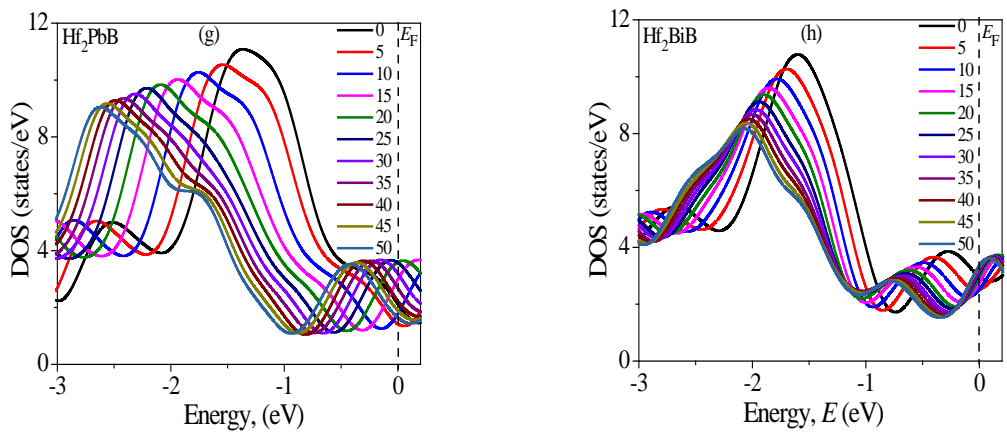

**Figs. 3:** Total density of states of (g) Hf₂PbB and (h) Hf₂BiB at different (0-50 GPa) pressures.

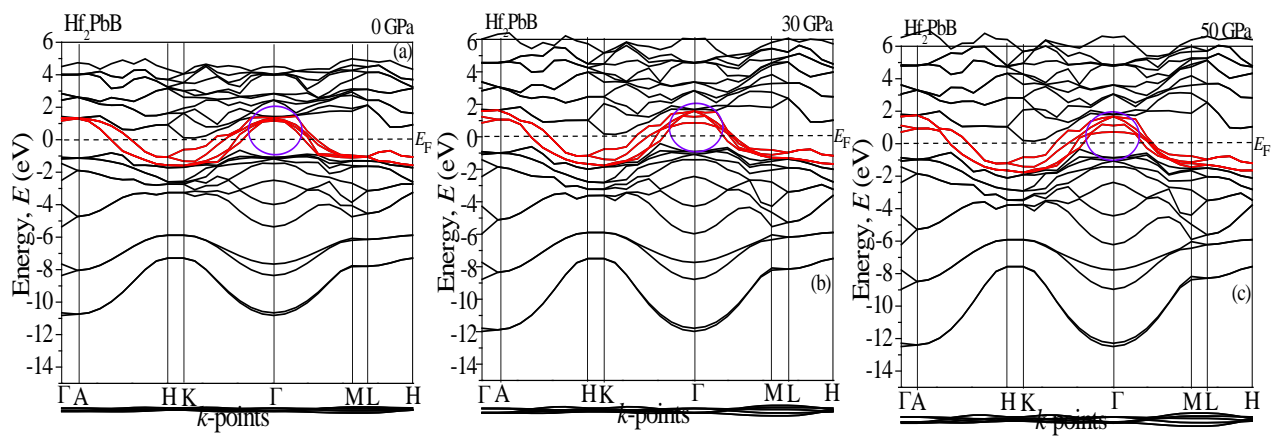

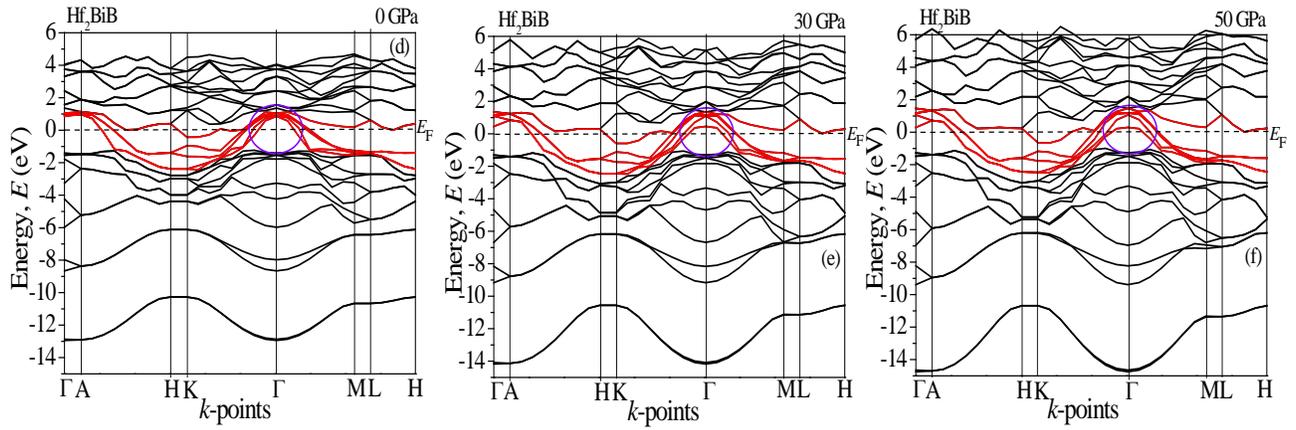

**Figs. 4:** Electronic band structure of Hf$_2$PbB (a, b and c) and Hf$_2$BiB (d, e and f) at 0 GPa, 30 GPa, and 50 GPa pressure.

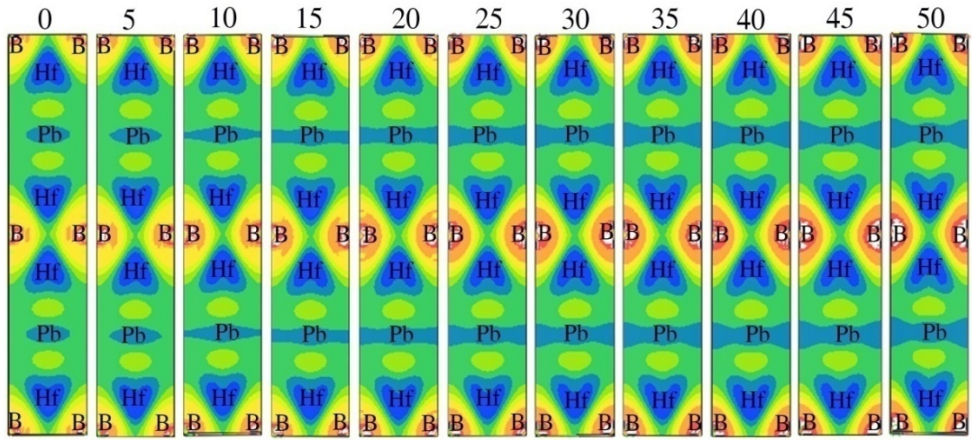

**Fig. 5(a):** Charge density mapping of Hf$_2$PbB phase at different (0-50 GPa) pressures.

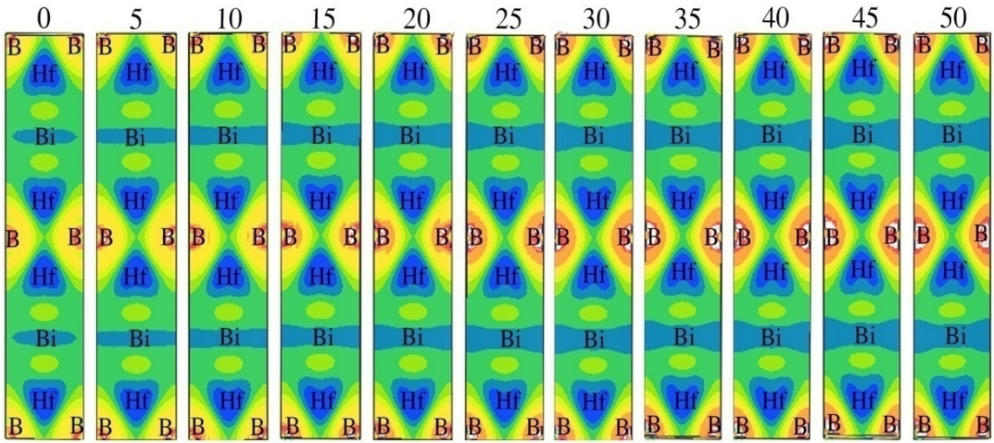

**Fig. 5(b):** Charge density mapping of Hf$_2$BiB phase at different (0-50 GPa) pressures.

### 3.3 Mechanical properties

Elastic constants can characterize the mechanical behavior of solids. The calculated elastic constants and moduli of the titled borides are shown in Table 2 within the pressure range (0–50 GPa). At 0 GPa, the elastic constants are consistent with the reported values [62]. Moreover, to become mechanically stable, the titled borides should satisfy the Born's stability criteria [63]. The conditions for the mechanical stability of the hexagonal system under pressure are as follows: $C_{44}' > 0$, $C_{11}' > |C_{12}'|$, $C_{33}'(C_{11}' + C_{12}') > 2C_{13}'^2$ where $C_{ii}' = C_{ii} - P (i = 1, 3)$ and $C_{1i}' = C_{1i} - P (i = 2, 3)$. The elastic constants of the studied borides satisfied the mechanical stability criteria within the examined pressure range. Figs. 6(a, b) shows the variation of elastic constants under applied pressure. As shown in the figure, the elastic constants are increased with an increase in pressure. It has been observed that $C_{11} > C_{33}$ up to 20 GPa, and after that, the increasing rate of $C_{33}$ is higher than that of $C_{11}$ for $Hf_2PbB$ phase. The increasing rate of $C_{33}$ dominates other elastic constants for $Hf_2BiB$ phase; $C_{66}$ increases slowly for both phase. Other elastic constants for the titled compounds increase monotonically. Elastic constants $C_{11}$ and $C_{33}$ measure the elastic response to uniaxial stress, while $C_{44}$ yields due to the shear stress. The unequal values of $C_{11}$ and $C_{33}$ indicate that the borides exhibit the anisotropic nature in the bonding strength. The obtained values of the $C_{ij}$ can be compared with those of other 211 MAX phase borides. At 0 GPa, the obtained $C_{11}$ of $Hf_2AB$ (A = Pb, Bi) are higher than those of $M_2AlB$ (M= Ti, Zr, Hf), but lower than those of $M_2AlB$ (M= V, Nb, Ta, Cr, and Mo) [25]. Whereas, the $C_{33}$ of $Hf_2AB$ (A = Pb, Bi) are higher than those of $M_2AlB$ (M=Zr, Hf), but lower than those of $M_2AlB$ (M= Ti, V, Nb, Ta, Cr, and Mo) [25]. Another important parameter, $C_{33}$, which is also higher for the titled compounds compared to those of $M_2AlB$ (M= Ti, Zr, Hf), but lower than those of $M_2AlB$ (M= V, Nb, Ta, Cr, and Mo). The elastic moduli of materials such as Young's modulus ($Y$), shear modulus ($G$), and bulk modulus ($B$) can be calculated from elastic stiffness constants using the formalism found elsewhere [64-66]. The values of $B$, $G$, and $Y$ also comparable to other 211 MAX phase borides like $C_{ij}$ [25]. These parameters are not directly related to the hardness of solids, but the larger value of these parameters corresponds to the harder materials. The calculated elastic moduli of the $Hf_2AB$ (A = Pb, Bi) borides show consistency with the reported values at 0 GPa pressure [62]. Table 2 shows a higher Young's modulus than other parameters ($B$, $G$); the ranges are 192-387 and 183-403 GPa for the $Hf_2PbB$ and $Hf_2BiB$ phase, respectively. The bulk modulus of $Hf_2PbB$ are in the range of 108-309 GPa within the studied pressure range, which is lower than those of $Hf_2BiB$ (115-323 GPa), indicating that $Hf_2BiB$ is harder than the $Hf_2PbB$. Fig. 6(c, d) shows the pressure effect on the elastic moduli. The increasing rate of shear modulus ($G$) is smaller than the bulk modulus ($B$) and Young's modulus ($Y$), indicating that shear deformation can easily occur. As seen in Table 2 and Fig. 6 (a-d), the elastic constants and moduli of the studied compounds are much higher at 50 GPa compared to those at 0 GPa. The inter-atomic distance in the lattice decreases with the hydrostatic pressure, so the density of bond energies increases. As a result, the elastic moduli are increased, and the lattice becomes harder than that at lower pressures. Similar increment of the elastic constants and moduli with hydrostatic pressure for other MAX systems ($Zr_2PbC$ [50], $Hf_2AB_2$ (A = In, Sn) [51]) has been reported. Increasing these parameters with an increase in pressure notifies that the titled borides to become harder.

Poisson's ratio ($v$) and Pugh ratio ($G/B$) help to understand the brittle or ductile nature of the MAX phase. The condition of brittleness or ductility of the materials is proposed by Frantsevich *et al.* [67]. He proposed that if Poisson's ratio is less than 0.26, it shows brittle nature; otherwise shows ductile nature. Fig. 6(e) shows the change in Poisson's ratio of the titled compounds. These MAX phases show a brittle nature at 0 GPa pressure which is in good agreement with other studies phases [68]. As seen from the figure, after 15 GPa and 25 GPa pressure, $Hf_2BiB$ and $Hf_2PbB$ behave in ductile nature. After 35 GPa, the increasing rate of both compounds is the same. Another indicator of ductile or brittle nature is Pugh's ratio ($G/B$) [69, 70]. If the condition $G/B < 0.571$ or $B/G > 1.75$ satisfied, the materials behave as ductile and otherwise brittle. Fig. 6(f) is drawn to understand the pressure effect on the Pugh ratio. The brittle to ductile transition is observed at pressures like $v$ for both studied compounds.

**Table 2:** The elastic constants (in GPa), elastic moduli (in GPa), Poisson's ratio, Pugh ratio of $Hf_2AB$ (A = Pb, Bi) at different pressures.

| $Hf_2AB$ | Pressure (GPa) | $C_{11}$ | $C_{12}$ | $C_{13}$ | $C_{33}$ | $C_{44}$ | $B$ | $G$ | $Y$ | $v$ | $G/B$ |
|---|---|---|---|---|---|---|---|---|---|---|---|
| | 00 | 226 | 59 | 51 | 206 | 75 | 108 | 80 | 192 | 0.20 | 0.74 |
| | 05 | 257 | 72 | 69 | 246 | 89 | 131 | 91 | 222 | 0.22 | 0.69 |
| | 10 | 288 | 86 | 87 | 283 | 103 | 153 | 101 | 248 | 0.23 | 0.66 |
| | 15 | 318 | 103 | 105 | 315 | 113 | 175 | 109 | 271 | 0.24 | 0.62 |
| | 20 | 345 | 117 | 123 | 349 | 126 | 196 | 118 | 295 | 0.25 | 0.60 |
| $Hf_2PbB$ | 25 | 371 | 136 | 143 | 379 | 132 | 218 | 123 | 311 | 0.26 | 0.56 |

| Phase | | | | | | | | | | |
|---|---|---|---|---|---|---|---|---|---|---|
| | 30 | 399 | 153 | 161 | 408 | 144 | 239 | 130 | 330 | 0.27 | 0.54 |
| | 35 | 416 | 166 | 175 | 434 | 152 | 255 | 135 | 344 | 0.28 | 0.53 |
| | 40 | 427 | 187 | 190 | 462 | 171 | 272 | 140 | 358 | 0.28 | 0.51 |
| | 45 | 449 | 204 | 211 | 497 | 177 | 294 | 145 | 371 | 0.29 | 0.49 |
| | 50 | 469 | 223 | 222 | 513 | 191 | 309 | 150 | 387 | 0.29 | 0.49 |
| | 00 | 206 | 67 | 70 | 212 | 82 | 115 | 74 | 183 | 0.24 | 0.64 |
| | 05 | 235 | 85 | 92 | 255 | 92 | 140 | 82 | 206 | 0.25 | 0.59 |
| | 10 | 272 | 97 | 114 | 310 | 115 | 166 | 98 | 246 | 0.25 | 0.59 |
| | 15 | 298 | 112 | 130 | 280 | 128 | 180 | 101 | 255 | 0.26 | 0.56 |
| | 20 | 318 | 126 | 149 | 377 | 138 | 206 | 112 | 284 | 0.27 | 0.54 |
| Hf$_2$BiB | 25 | 344 | 144 | 166 | 393 | 150 | 225 | 118 | 301 | 0.28 | 0.52 |
| | 30 | 363 | 159 | 181 | 462 | 165 | 246 | 128 | 327 | 0.28 | 0.52 |
| | 35 | 395 | 167 | 190 | 434 | 175 | 257 | 135 | 337 | 0.28 | 0.52 |
| | 40 | 424 | 183 | 221 | 497 | 194 | 287 | 146 | 374 | 0.28 | 0.51 |
| | 45 | 428 | 197 | 231 | 500 | 204 | 295 | 145 | 395 | 0.29 | 0.49 |
| | 50 | 469 | 213 | 257 | 532 | 218 | 323 | 156 | 403 | 0.29 | 0.48 |

The Vickers hardness ($H_v$) of Hf$_2$AB (A = Pb, Bi) was also calculated by using Gou *et al.*'s formula [71] for partial metallic solids. As seen from Table 3, the Vickers hardness increases with the increase in pressure, meaning the bonds become stronger by an applied pressure owing to shortened bond lengths. Within the studied pressure range, the increase of elastic constant $C_{11}$ is found to be 107% for Hf$_2$PbB and 127% for Hf$_2$BiB from 0 to 50 GPa. Similarly, $C_{12}$ increases by 278% and 218%, $C_{13}$ by 335% and 267%, $C_{33}$ by 149% and 151%, and $C_{44}$ by 154% and 166% for Hf$_2$PbB and Hf$_2$BiB compounds, respectively from 0 to 50 GPa.

**Table 3:** Calculated Mulliken bond length $d^{\mu}$, bond overlap population $P^{\mu}$, metallic population $P^{\mu'}$, bond volume $v_b^{\mu}$, bond hardness $H_v^{\mu}$ of $\mu$-type bond and Vickers hardness $H_v$ of Hf$_2$AB (A = Pb, Bi) at different pressures.

| Phase | Pressure (GPa) | Bond | $d^{\mu}$ (Å) | $P^{\mu}$ | $P^{\mu'}$ | $v_b^{\mu}$ | $H_V^{\mu}$ (GPa) | $H_V$ (GPa) |
|---|---|---|---|---|---|---|---|---|
| | 0 | B-Hf | 2.373 | 1.68 | 0.102 | 39.365 | 2.562 | 2.562 |
| | 30 | B-Hf | 2.271 | 1.67 | 0.126 | 6.645 | 48.633 | 3.840 |
| Hf$_2$PbB | | Hf-Hf | 4.524 | 0.15 | | 52.523 | 0.023 | |
| | 50 | B-Hf | 2.227 | 1.64 | 0.138 | 6.332 | 51.269 | 5.995 |
| | | Hf-Hf | 4.391 | 0.21 | | 48.532 | 0.081 | |
| | 0 | B-Hf | 2.388 | 1.73 | 0.017 | 8.087 | 38.888 | 3.681 |
| | | Hf-Hf | 4.694 | 0.06 | | 61.385 | 0.033 | |
| | 30 | B-Hf | 2.282 | 1.69 | 0.017 | 7.262 | 45.445 | 6.792 |
| Hf$_2$BiB | | Hf-Hf | 4.364 | 0.16 | | 50.753 | 0.151 | |
| | 50 | B-Hf | 2.236 | 1.67 | 0.014 | 6.858 | 49.476 | 8.728 |
| | | Hf-Hf | 4.250 | 0.24 | | 47.056 | 0.271 | |

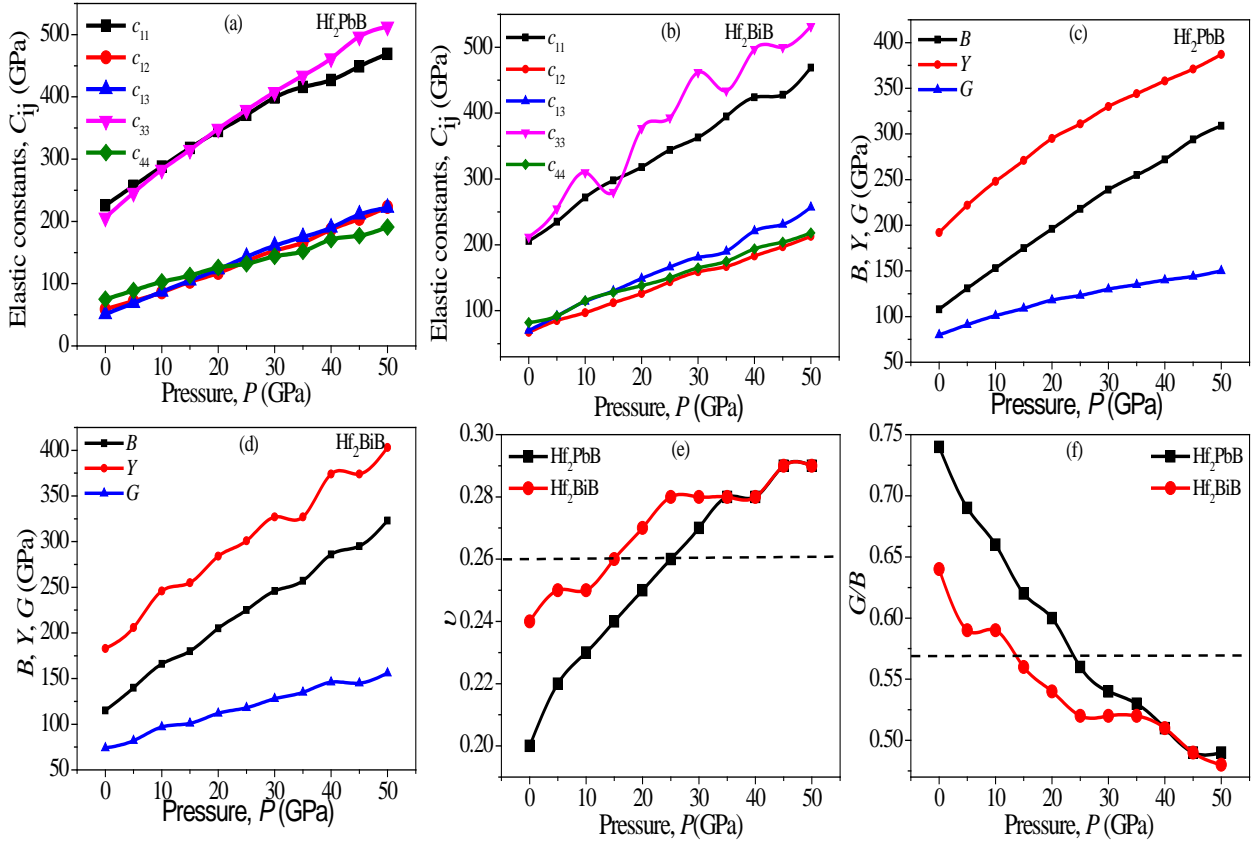

**Figs. 6:** Variation of elastic constants of (a) Hf$_2$PbB and (b) Hf$_2$AB; elastic moduli of (c) Hf$_2$PbB and (d) Hf$_2$AB; (e) Poisson's ratio ($v$) and (f) Pugh ratio ($G/B$) of Hf$_2$AB (A = Pb, Bi) phase as a function of pressure.

## 3.4 Thermal properties

MAX phases show good mechanical properties at high temperatures, which can be used in high-temperature technology. Debye temperature ($\Theta_D$), minimum thermal conductivity ($K_{min}$), Grüneisen parameter ($\gamma$), and melting temperature ($T_m$) have been studied to assess the appropriateness of the titled borides in this section.

Debye temperature represents the highest normal mode of vibration of the crystal's frequency. Debye temperature and elastic properties are associated with thermodynamic properties such as phonons, specific heat, melting temperature, and thermal expansion of materials. A Higher Debye temperature indicates that the corresponding phase is harder [72, 73]. Moreover, elastic moduli and constants will be higher for the harder solids. Average sound velocity ($v_m$) is correlated with the crystal shear and bulk modulus, and Debye temperature has been calculated using the value of $v_m$. As the Vickers hardness of the Bi-containing boride is more than that of Pb-containing boride, it is expected that the Debye temperature of the Bi-containing boride is also higher than that of Pb-containing boride. But one can see from Table 4 that this temperature is slightly higher for $Hf_2PbB$ compared to $Hf_2BiB$ at 0 GPa. It is assumed that the atomic mass of $Hf_2BiB$ borides is higher than the $Hf_2PbB$ boride, that's why the higher density of Bi borides results in a lower $\Theta_D$. Similar results are also reported for other phases [74]. In addition, $\Theta_D$ is also close to that (342 K) of the $Hf_2AlB$ boride [25]. As seen from Fig. 7(a), Debye temperature increases with an increase in pressure that notifies the increase of lattice vibrational limit under the studied pressure range (0-50) GPa. The increase of Debye temperature with pressure certifies the enhancement of hardness of the titled compounds with rising pressure. The calculated Debye temperature, minimum thermal conductivity, Grüneisen parameter, and melting temperature have been listed in Table 4.

The temperature responsible for the phase transition from solid to liquid is known as melting temperature. An empirical method is developed by Fine *et al*. [75] to predict the melting temperature. A close relationship between melting temperature and Young's modulus is observed. Generally, solids with a higher Young's modulus exhibit a higher melting temperature. As a reference, the $Hf_3PB_4$ phase with Young's modulus 426 GPa shows a higher melting temperature (2282 K) than the titled borides [74]. From Fig. 7(b), one can see that the melting temperature is increased with an increase in pressure, which suggest that these compounds can be used as structural materials in extreme environments, i.e., in high-temperature technology at elevated pressure.

**Table 4:** The computed density ($\rho$), mean sound velocities ($V_m$), Debye temperature ($\Theta_D$), melting temperature ($T_m$), minimum thermal conductivity ($K_{min}$), and Grüneisen parameter ($\gamma$) of $Hf_2AB$ (A = Pb, Bi) borides at different pressures.

| $Hf_2AB$ | Pressure | $\rho(g/cm^3)$ | $v_m(m/s)$ | $\Theta_D(K)$ | $T_m(K)$ | $K_{min}$ (Wm$^{-1}$K$^{-1}$) | $\gamma$ |
|---|---|---|---|---|---|---|---|
| | 0 | 12.13 | 2836 | 313 | 1341 | 0.54 | 1.29 |
| | 5 | 12.64 | 2968 | 332 | 1494 | 0.58 | 1.37 |
| | 10 | 13.09 | 3077 | 348 | 1587 | 0.61 | 1.41 |
| | 15 | 13.49 | 3152 | 360 | 1780 | 0.64 | 1.45 |
| | 20 | 13.86 | 3239 | 374 | 1912 | 0.67 | 1.50 |
| | 25 | 14.20 | 3272 | 380 | 2035 | 0.69 | 1.55 |
| | 30 | 14.51 | 3331 | 390 | 2163 | 0.71 | 1.60 |
| $Hf_2PbB$ | 35 | 14.81 | 3362 | 397 | 2253 | 0.73 | 1.66 |

| | | | | | | |
|---|---|---|---|---|---|---|
| | 40 | 15.08 | 3395 | 403 | 2328 | 0.74 | 1.66 |
| | 45 | 15.35 | 3417 | 408 | 2446 | 0.76 | 1.71 |
| | 50 | 15.60 | 3459 | 415 | 2530 | 0.77 | 1.71 |
| | 0 | 12.35 | 2713 | 301 | 1290 | 0.52 | 1.45 |
| | 5 | 12.84 | 2807 | 315 | 1495 | 0.55 | 1.50 |
| | 10 | 13.27 | 3018 | 343 | 1635 | 0.61 | 1.50 |
| | 15 | 13.67 | 3023 | 347 | 1668 | 0.62 | 1.55 |
| | 20 | 14.03 | 3144 | 364 | 1873 | 0.65 | 1.60 |
| | 25 | 14.36 | 3192 | 372 | 1975 | 0.67 | 1.66 |
| | 30 | 14.67 | 3290 | 387 | 2136 | 0.71 | 1.66 |
| Hf$_2$BiB | 35 | 14.96 | 3345 | 396 | 2190 | 0.73 | 1.66 |
| | 40 | 15.24 | 3450 | 411 | 2371 | 0.76 | 1.66 |
| | 45 | 15.50 | 3411 | 408 | 2388 | 0.76 | 1.71 |
| | 50 | 15.76 | 3511 | 422 | 2559 | 0.79 | 1.71 |

Study of minimum thermal conductivity ($k_{min}$) is important for choosing materials usable in high-temperature applications. The acoustic wave velocity of phonons and phonon's density of states are closely related to this temperature, and this temperature becomes constant at high temperatures. The value of $k_{min}$ of 1.25 $Wm^{-1}K^{-1}$ has been considered a based value for thermal barrier coating application. The minimum thermal conductivity of selected compounds has been calculated [76] and is 0.54 and 0.52 $Wm^{-1}K^{-1}$ at 0 GPa for Hf$_2$PbB and Hf$_2$BiB, respectively, which lower than those of M$_2$AlB (M= Ti, Zr, Hf, V, Nb, Ta, Cr, and Mo) [25]. These values are close to other MAX phases so that the titled borides can be used as an alternative to the reported phases [77, 78]. It increases linearly with increasing pressure [Fig. 7(c)] but remained lesser than that of the limiting value; hence, the possibility of choosing them as coating materials has been revealed up to considered pressure range.

The lattice dynamics of crystalline solids generate anharmonic effects that can be understood from the Grüneisen parameter ($\gamma$). It is associated with crystal-specific heat, thermal expansion coefficient, volume, etc. This value is closely related to Poisson's ratio. The values of $\gamma$ are found to be .85 to 3.53 for Poisson's ratio of 0.05 to 0.46 of the polycrystalline solids [79]. The computed values of this parameter fall within the condition mentioned above [Table 4]. This parameter is increased with pressure, that's why the anharmonicity of the Hf$_2$AB (A = Pb, Bi) has been enhanced [Fig. 7(d)]. However, comparing the values of $\Theta_{D,}$ $K_{min}$, $T_m$, and $\gamma$ of Hf$_2$AB with those of Y$_4$Al$_2$O$_9$ [80, 81], it is confirmed that the titled borides are also possible candidates for high-temperature applications.

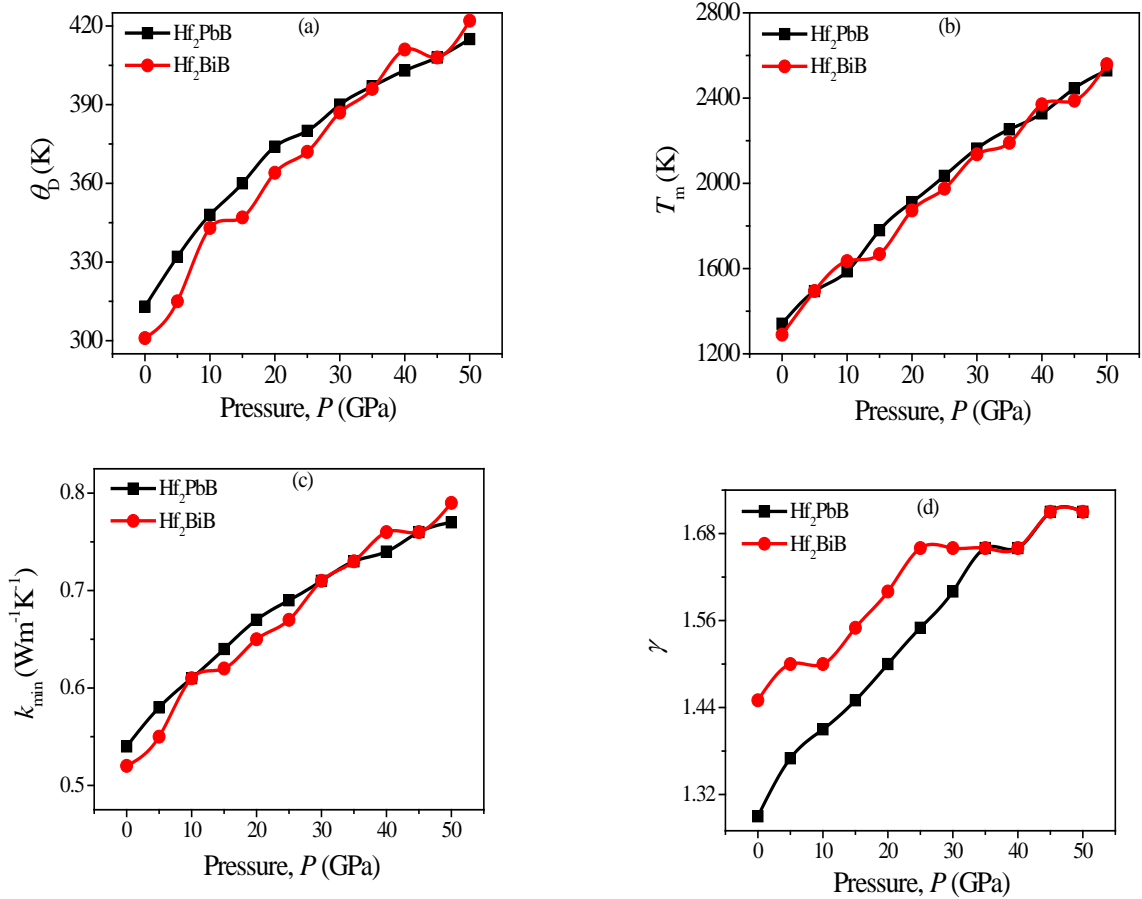

**Figs. 7:** Variation of (a) Debye temperature($\theta_D$), (b) Melting temperature ($T_m$), (c) Minimum thermal conductivity ($K_{min}$), and (d) Grüneisen parameter ($\gamma$) of Hf$_2$AB (A = Pb, Bi) borides as a function of pressure.

### 3.5 Optical properties

The imaginary part $\varepsilon_2(\omega)$ of the dielectric function measures the loss of photon energy due to absorption, and another part [real part $\varepsilon_1(\omega)$] contributes to polarization. Fig. 8(a, b) illustrates the computed optical constants using the equation found elsewhere for both compounds [82, 83]. The obtained optical properties are compared (distinctive values, figures are not shown) with those of widely studied 211 MAX phases those of Ti$_2$AlC and Ti$_2$AlN measured by Haddad et al. [82] and 211 MAX phase borides M$_2$SB (M= Zr, Hf, Nb) [33]. As seen Fig. 8, the real and imaginary curves for both polarization directions of these compounds approach zero between 13.5 eV to 16 eV. It is noted that the real part comes from below, and the imaginary part comes from above. Haddad et al. [82] also obtained similar pattern for $Ti_2AlC$ and $Ti_2AlN$. The large negative values of the real part assure the Drude-like nature of considered compounds, confirming the metallic nature of them, like other MAX phases [3, 33, 78]. The major peaks are observed in the range of 0.6-0.9 eV and 0.5-0.9 eV for [100] direction, while peaks are observed in the range of 0.7-1.4 eV and 0.8-1.7 eV for [001] direction, respectively. In addition, the value of $\varepsilon_1(\omega)$ is higher along the [100] direction than that of the [001] direction for both phase, confirming the optical anisotropy. The $\varepsilon_2(\omega)$ peaks indicate electron excitation upon the incident photon on solids and the highest peak of this value is responsible for the highest absorption. The $\varepsilon_2(\omega)$ exhibits the sharp peaks at around 4.5-6 eV and 5 eV for the $Hf_2PbB$ and $Hf_2BiB$, respectively. It is noticed that the peaks shift to the higher energy range with increasing pressure. The nature of $\mathcal{E}_1$ and $\mathcal{E}_2$ spectra of the studied compounds is similar to those of the most studied MAX phase $Ti_3SiC_2$ [83].

The extinction coefficient ($k$) and refractive index ($n$) are shown in Fig. 8(c, d). The velocity of light in the materials is determined by $n$, whereas $k$ measures the attenuation of photon. Sharp peaks are observed for both borides owing to their intra-band transition. The static value $n(0)$ is the maximum refractive index, which is further decreased with increasing photon energy, like $M_2SB$ (M= Zr, Hf, Nb) [33]. The maximum losses ($k$) are seen at around 5 eV along the [001] direction. Refractive indices increase due to the higher density of compounds, whereas the extinction coefficient is decreased. When a photon energy incident on materials, some portions of the energy penetrate materials and some are absorbed by it, whereas the rest are reflected.

The absorption coefficient ($\alpha$) measures the magnitude of how amounts enter into the materials during propagation. The highest value of this parameter indicates maximum absorption under a particular energy range. One can understand the characters of the absorption coefficient from Fig. 8(e). As seen from the Fig 8 (e), the absorption started at 0 eV owing to its metallic nature. The spectra are higher within the ultraviolet region for both compounds than in other regions. For [001] direction, the maximum absorption value for $Hf_2PbB$ and $Hf_2BiB$ compounds lays in the range of 5.50-6.50 eV and 5.50-6.60 eV, respectively. The highest peaks are found around 6.50-7.50 eV and 6-7 eV in the [100] polarization for $Hf_2PbB$ and $Hf_2BiB$, respectively. For $M_2SB$ (M= Zr, Hf, Nb) [33], the maximum is observed in the range of 7.0 to 9.0 eV. It is seen that the absorption coefficient along the [001] direction is greater than that of [100] direction, confirming their anisotropic nature. The spectra are shifted to the higher energy range due to the pressure effect and show a maximum absorption coefficient at 50 GPa pressure for both compounds. The high absorption nature at the UV region also messages that these are competing phases for optoelectronic devices and can be used in industrial applications.

The photoconductivity ($\sigma$) of the studied compounds is shown in Fig. 8(f) for two directions. It measures the variation of conductivity after the incident of photons. The photoconductivity reaches saturation at the highest absorption. In addition, peaks value for the related materials has been observed concerning different energies and polarization. The peaks are found in the range of 4.60 to 5.50 eV and 4.75 o 5 eV in the [001] direction, nearby position of the absorption spectra, for $Hf_2PbB$ and $Hf_2BiB$, respectively. The peaks of photoconductivity are also appeared at slightly low energy side for $M_2SB$ (M= Zr, Hf, Nb) [33].

The reflectivity at different pressure is shown in Fig. 8(g). These graphs show photon energy's function for two polarization directions, and the change of reflectivity has been observed. At the infrared region, the spectra of reflection have been decreased. It is worth noting that in the visible region, the spectra of the $Hf_2BiB$ phase have been decreased, whereas $Hf_2PbB$ is increased. At the UV region, the reflectivity of both phase has been increased significantly. The reflectivity has drastically reduced after 14.40 to 14.70 eV and 14.50 to 14.98 eV for $Hf_2PbB$ and $Hf_2BiB$, respectively [in the [001] direction]. A sudden reduction of the reflectivity for $M_2SB$ (M= Zr, Hf, Nb) is also shown around 15.0 eV incident photon energy [33]. The initial reflectivity is found to be 45%, and the highest values are 100% and 97% of these compounds, respectively. Hence, these are probable candidates as coating materials [83, 84]. Although the pressure effect on the reflectance has been small and anisotropic effect is pronounced after 10 eV for both compounds.

Fig. 8(h) shows the calculated loss function [$L(\omega)$] of $Hf_2AB$ (A = Pb, Bi) compounds, which is used to evaluate the loss of energy of the electrons when traversing within the solids. Thus, the spectra exhibit zero value corresponding to the no loss of electron's energy. The peak observed in the spectra is termed as the plasma frequency $\omega_p$ of the material, which is the characteristic frequency at which the $\varepsilon_1(\omega)$ crosses the zero value from below when the $\varepsilon_2(\omega) < 1$. The material exhibits a dielectric response after $\omega_p$. The $\omega_p$ of $Hf_2PbB$ and $Hf_2BiB$ for [100] polarization direction are observed in the range of 13.6 eV to 14.05 eV and 13.8 eV to 13.9 eV, respectively at different pressures. The $\omega_p$ is observed at 14.26, 14.9 and 15.05 eV for Zr$_2$SB, Hf$_2$SB and Nb$_2$SB, respectively. The measured $\omega_p$ of Ti$_2$AlC and Ti$_2$AlN is around 17 eV [82], which is comparable to the present results. Moreover, there is no peak in the energy range 0-12.5 eV due to the large value of $\varepsilon_2(\omega)$. The peaks shift to the higher energy range as pressure is increased.

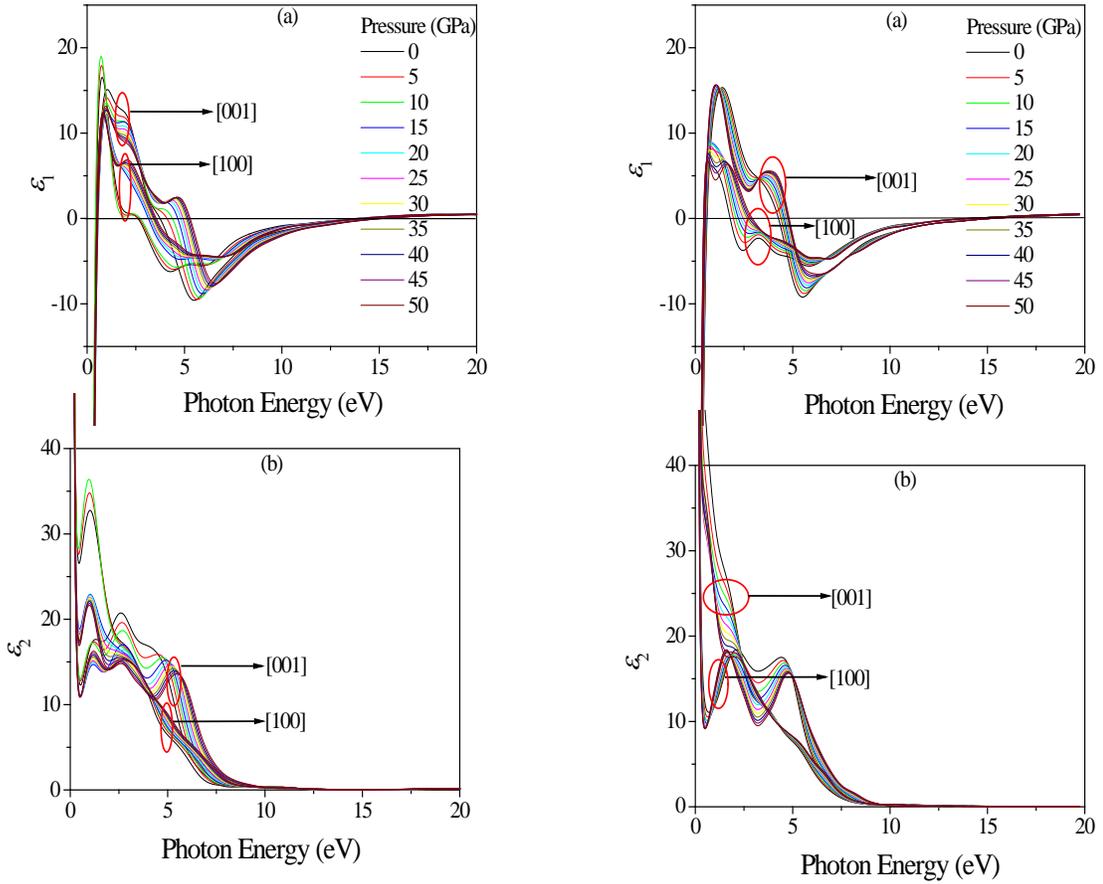

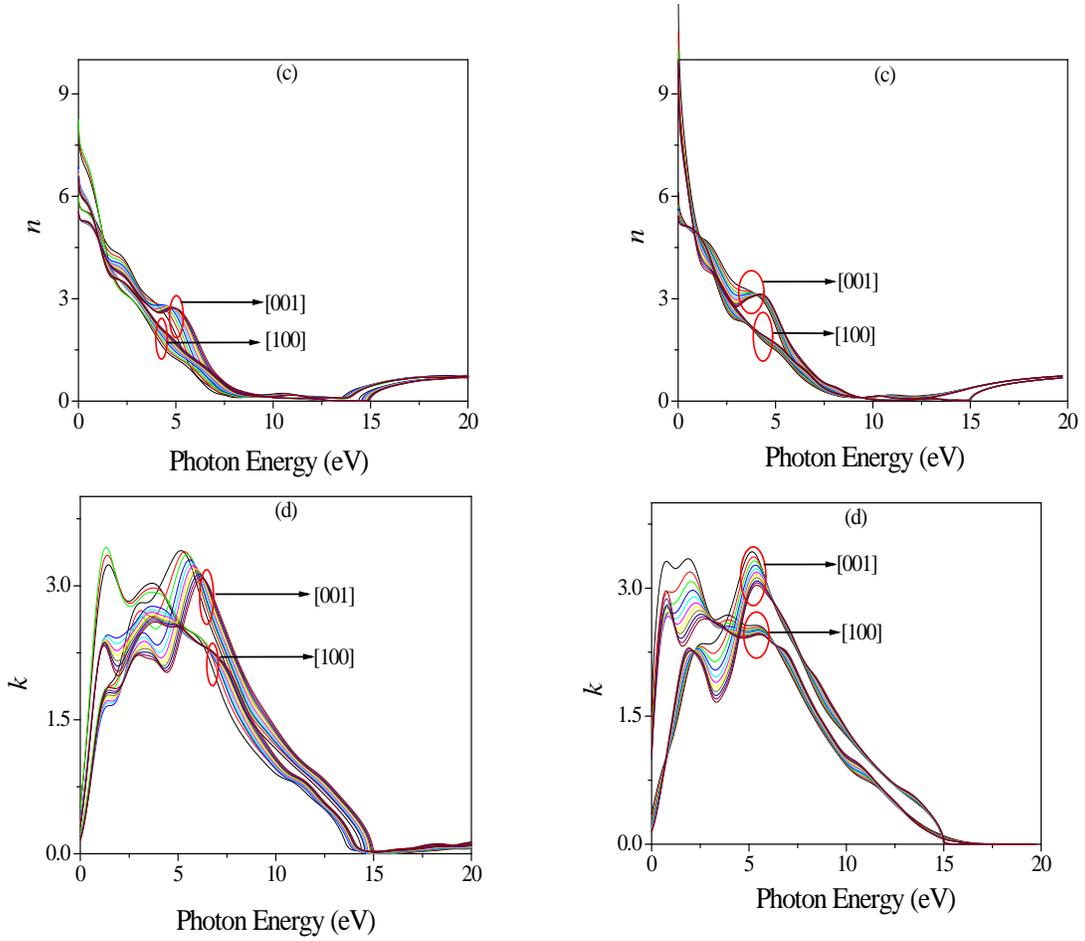

**Fig. 8:** (a) Real part of dielectric function, $\varepsilon_1(\omega)$, (b) imaginary part of dielectric function, $\varepsilon_2(\omega)$, (c) refractive index ($n$), and (d) extinction coefficient ($k$) of Hf$_2$PbB (Left panel) and Hf$_2$BiB (Right panel) phase at different pressures.

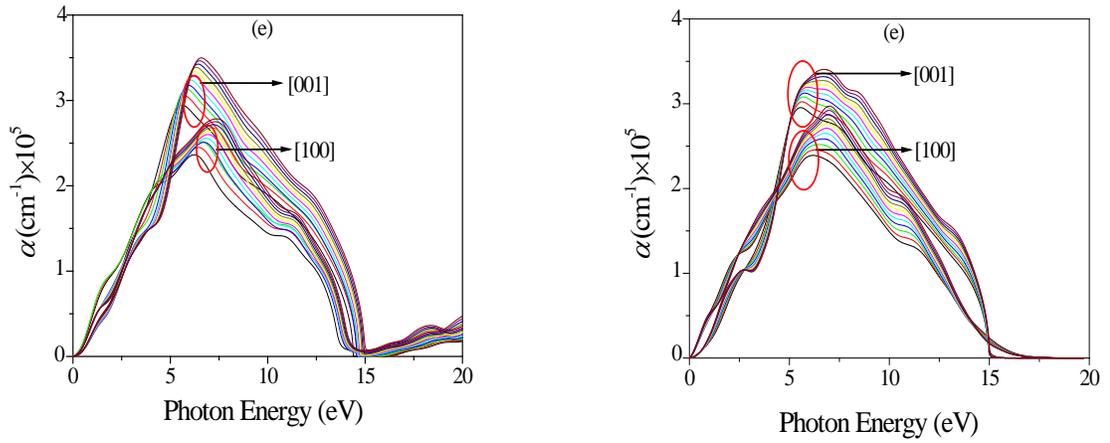

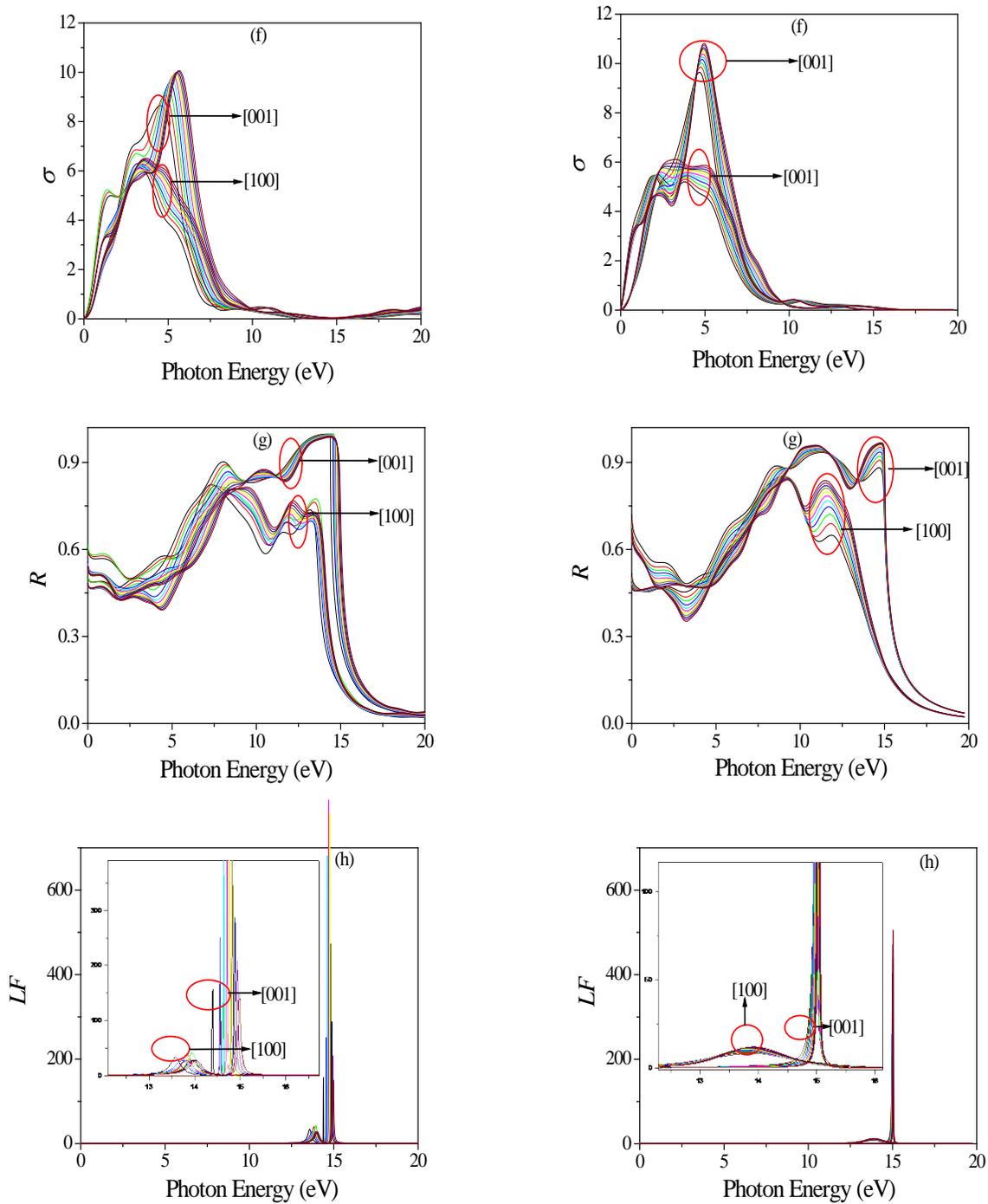

**Fig. 8:** (e) Absorption (α), (f) conductivity (σ), (g) reflectivity (*R*), and (h) loss function (*LF*) of Hf$_2$PbB (Left panel) and Hf$_2$BiB (Right panel) phase at different pressures.

## 4. Conclusions

The investigation of $Hf_2AB$ (A = Pb, Bi) MAX phase borides under pressure has been performed via DFT simulations. Normalized lattice parameters ($a/a_0$, $c/c_0$, and $v/v_0$) decreases with pressure while an internal parameter ($Z_M$) and hexagonal ratio ($c/a$) increases. The $O_r$ decreases with increasing pressure for both phase, while the $P_r$ follows a reverse trend. Electronic band structure and density of states certify the metallic nature of the titled compounds wherein the anisotropic electronic conductivity is revealed. The DOS of $Hf_2PbB$ is higher than that of $Hf_2BiB$ with a dominant contribution from the Hf-$5d$ orbital for both compounds. The charge density variation is also seen with increased pressure for both compounds, which certifies the variation of bond strength and hardness with pressure. The pressure-dependent stability criteria and phonon dispersion curves confirm the stability of titled compounds up to the studied pressure. The elastic constants increase with increasing pressure, as expected. The elastic moduli are also found to be increased with pressure. The increasing rate of shear modulus is lower than that of bulk and Young's modulus. The titled MAX phase borides show brittle characters at ambient pressure; a transition from brittle to ductile nature occurs for both with applying pressure. At 0 GPa, the Vickers hardness of $Hf_2BiB$ phase (2.85 GPa) is higher than that of the $Hf_2PbB$ (2.56 GPa) phase. The values of Debye temperature, melting temperature, and minimum thermal conductivity suggest their possible use in high-temperature technology such as TBC materials. The Debye temperature, melting temperature, minimum thermal conductivity, and Grüneisen parameter is also increased with the increase in pressure owing to the close relationship with elastic parameters. A minor pressure effect on the optical properties is observed. The results of dielectric constants, absorption coefficient, and photoconductivity agree with band structure results. The reflectivity spectra suggest that the studied borides are appropriate candidates for use as coating materials to protect from solar heating. The unequal values of the optical constants for two different directions indicate the optical anisotropy of the studied MAX compounds.

## Declaration of interests

The authors declare that they have no known competing financial interests or personal relationships that could have appeared to influence the work reported in this paper.

**Acknowledgments**

The authors are grateful to the Planning and Development, Chittagong University of Engineering & Technology (CUET), Chattogram-4349, Bangladesh, for providing the financial assistance for this work. A part of this work has been carried out at Advanced Computational Materials Research Laboratory (ACMRL), Department of Physics, Chittagong University of Engineering and Technology (CUET), Chattogram-4349, Bangladesh, which is developed by the aid of a grant (No. 21-378 RG/PHYS/AS_G-FR3240319526) from UNESCO-TWAS and the Swedish International Development Cooperation Agency (Sida). The views expressed herein do not necessarily represent those of UNESCO-TWAS, Sida or its Board of Governors.